\documentclass{article}
\pdfoutput=1

\usepackage{bm}
\usepackage{graphicx}
\usepackage{natbib}
\usepackage{hyperref}

\begin{document}

\title{Direct multi-scale reconstruction of velocity fields from measurements of particle tracks}

\author{D. H. Kelley and N. T. Ouellette \\
Department of Mechanical Engineering \\
Yale University, New Haven, CT 06520, USA \\
\href{http://leviathan.eng.yale.edu}{http://leviathan.eng.yale.edu} \\
\href{mailto:nicholas.ouellette@yale.edu}{nicholas.ouellette@yale.edu} }

\date{\today}

\maketitle

\begin{abstract}

We present a method for reconstructing two-dimensional velocity fields at specified length scales using observational data from tracer particles in a flow, without the need for interpolation or smoothing. The algorithm, adapted from techniques proposed for oceanography, involves a least-squares projection of the measurements onto a set of two-dimensional, incompressible basis modes with known length scales. Those modes are constructed from components of the velocity potential function, which accounts for inflow and outflow at the open boundaries of the measurement region; and components of the streamfunction, which accounts for the remainder of the flow. All calculations are evaluated at particle locations, without interpolation onto an arbitrary grid. Since the modes have a well-defined length scales, scale-local flow properties are available directly. The technique eliminates outlier particles automatically and reduces the apparent compressibility of the data. Moreover the technique can be used to produce spatial power spectra and to evaluate the spatial effects of open boundaries; it also holds promise for direct calculation of scale-to-scale transfer of enstrophy and energy. 

\end{abstract}

\section{Introduction}

Over the past two decades, particle-based video measurements have grown robust enough to become standard tools in experimental fluid mechanics. The powerful techniques of Particle Image Velocimetry (PIV) and its cousin Particle Tracking Velocimetry (PTV) are widely used to obtain Eulerian flow information \citep{Tropea:2007}, and under proper conditions can be very accurate. More general particle-tracking methods can be used to obtain Lagrangian information, allowing the measurement of, for example, the statistics of the velocity gradient along particle trajectories \citep{Luthi:2005}, the so-called Lagrangian Coherent Structures that contain information about the stretching of material lines \citep{Voth:2002,Mathur:2007}, and the motion of topological singularities in the flow \citep{Ouellette:2007}. 

All of these techniques involve approximating the flow velocity \emph{field}---a vector function of continuous arguments---from measurements at a discrete set of locations. The full field is then usually obtained by interpolation between the points, generally either linearly or using cubic splines. Such an approach is often reasonable, and is widely applied in computational fluid dynamics (CFD) models, where the equations are solved on a discrete mesh. Unlike in CFD, however, these experimental velocimetry methods have no control over the locations of the mesh points: velocities are known only at the locations of the particles, which are randomly positioned throughout the flow. In particular, experimentalists cannot employ techniques such as adaptive mesh refinement in regions where the flow changes rapidly in space. Simple interpolation, then, can introduce significant error into the resulting flow fields. To mitigate such effects, experimentalists often use some kind of smoothing filter on the data; the choice of filter type and parameters, however, can be more an art than a science. We seek instead a data-processing scheme that imposes only a well defined set of physical constraints on the measurements with as little averaging or smoothing as possible, so that the data itself determines the spatial resolution and coherence.

Here, we present an alternate methodology for reconstructing the velocity field that has a sound mathematical foundation and requires no \textit{ad hoc} smoothing. Our technique borrows tools employed in oceanography~\citep{Lynch:1989,Chu:2003,Lekien:2004}, where researchers have long faced the challenge of systematically constructing velocity field estimates from poorly sampled data. As we describe in detail below, we first choose a set of basis functions that are appropriate for the measured data, imposing physical constraints such as incompressibility on the form of these functions to control the properties of the resulting reconstructed velocity field. The data are then projected onto the basis functions in a least-square sense~\citep{Press:2007}; no additional smoothing is required. 

By combining the powerful techniques developed by oceanographers for handling sparse data sets with the high spatial resolution provided by modern optical flow diagnostics, our reconstruction method gives us access to remarkably clean and well controlled data sets. It also gives direct access to the scale-local field properties: since each basis function has a well-defined length scale, we can easily project onto a subset of those modes to construct spatially resolved fields with known spectral content (low-pass, high-pass, or band-pass). This type of information has recently been exploited to gain a detailed understanding of the spatial structures responsible for scale-to-scale enstrophy and energy transfer \citep{Rivera:2003,Bruneau:2007}, but is typically obtained using more complex and less direct methods such as filter-space techniques \citep{Rivera:2003} or wavelet filters \citep{Bruneau:2007}. 

We have developed and tested our reconstruction method using data from a quasi-two-dimensional, electromagnetically forced, shallow, stratified flow. In \S\ref{sec:experiment}, we describe the flow in detail and discuss the particle-tracking system we use to make measurements. In \S\ref{sec:reconstruction}, we introduce the reconstruction technique, including the theoretical background for the method and some details of its actual implementation. Of particular interest is the choice of basis modes, and we discuss the relative strengths and weaknesses of both Fourier modes and eigenmodes calculated numerically. In \S\ref{sec:structure}, we give examples of how our reconstruction scheme can be used to obtain spatially resolved scale-local measurements. Finally, in \S\ref{sec:conclusion} we summarize our results and give some directions for future work.

\section{Experimental Methods}
\label{sec:experiment}

\subsection{Flow}

\begin{figure}
\begin{center}
\includegraphics[width=\textwidth]{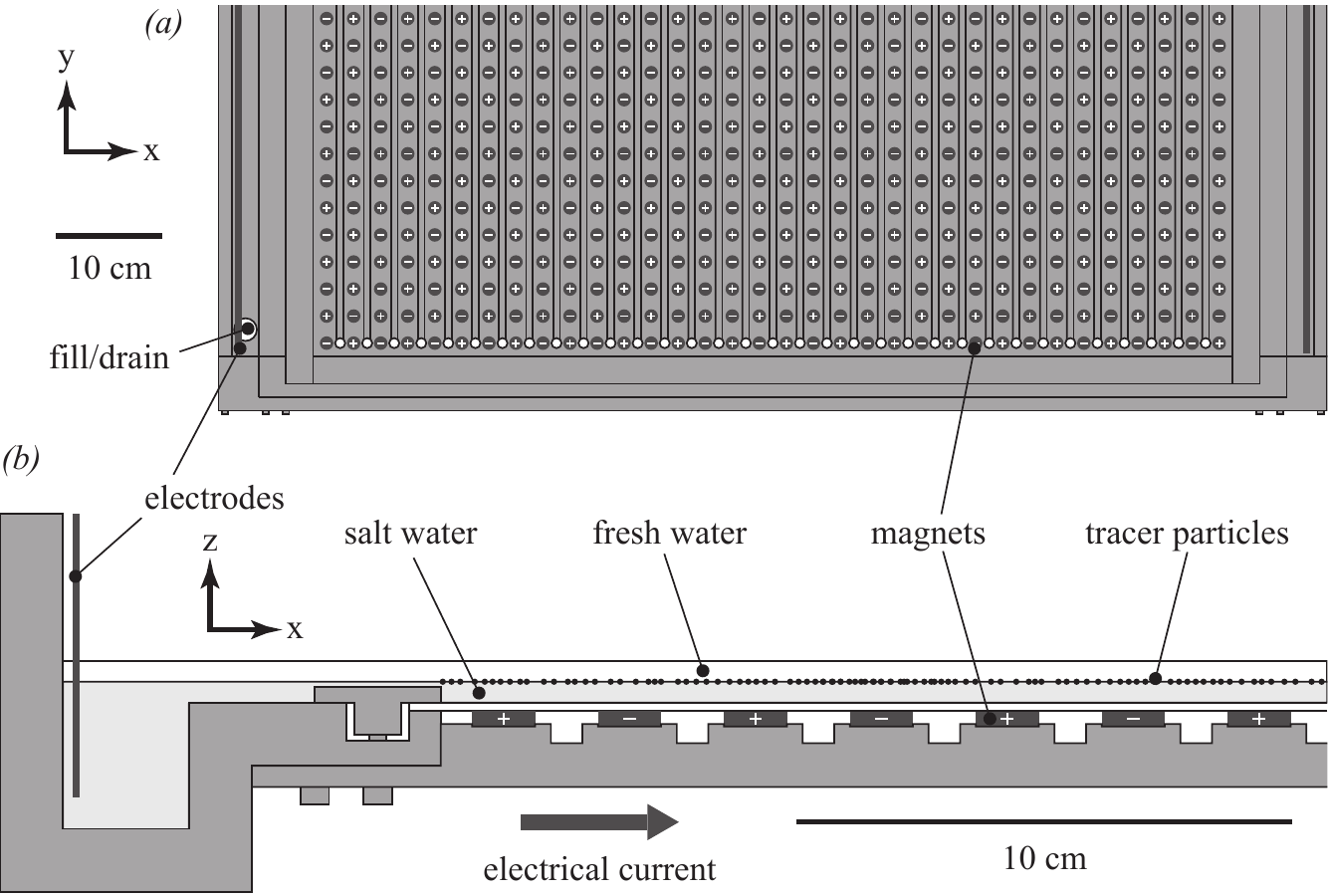}
\caption{\label{fig:apparatus} The experimental apparatus, as seen (a) empty from above, and (b) filled in cross-section. Both sketches are partial; the full test bed is square, with walls on all sides. Electrical current flows from left to right through the saltwater layer.}
\end{center}
\end{figure}

We have designed and constructed an experimental apparatus, sketched in figure~\ref{fig:apparatus}, intended to serve as a physical model of two-dimensional fluid flows. Essentially a shallow pan of stably-stratified salt water, it uses electromagnetic forcing and resembles previous devices~\citep{Tabeling:1991,Rothstein:1999,Solomon:2003,Clercx:2003,Rivera:2005,Rossi:2006}, though it is larger than many. Its test bed has a smooth, glass floor 86~cm x 86~cm x 3.2~mm thick. We coat the floor with a hydrophobic wax to reduce friction, cover it with deionized water, and then inject heavier salt water from below such that the stratification is gravitationally stable (denser salt water below lighter freshwater). During the experiments discussed below, the saltwater solution was NaCl in water, 16\% by mass, with a density of $\rho=1 116$~kg/m$^3$ (specific gravity 1.116) and a kinematic viscosity of $\nu=1.24 \times 10^{-6}$~m$^2$/s. 

\begin{figure}
\begin{center}
\includegraphics[width=\textwidth]{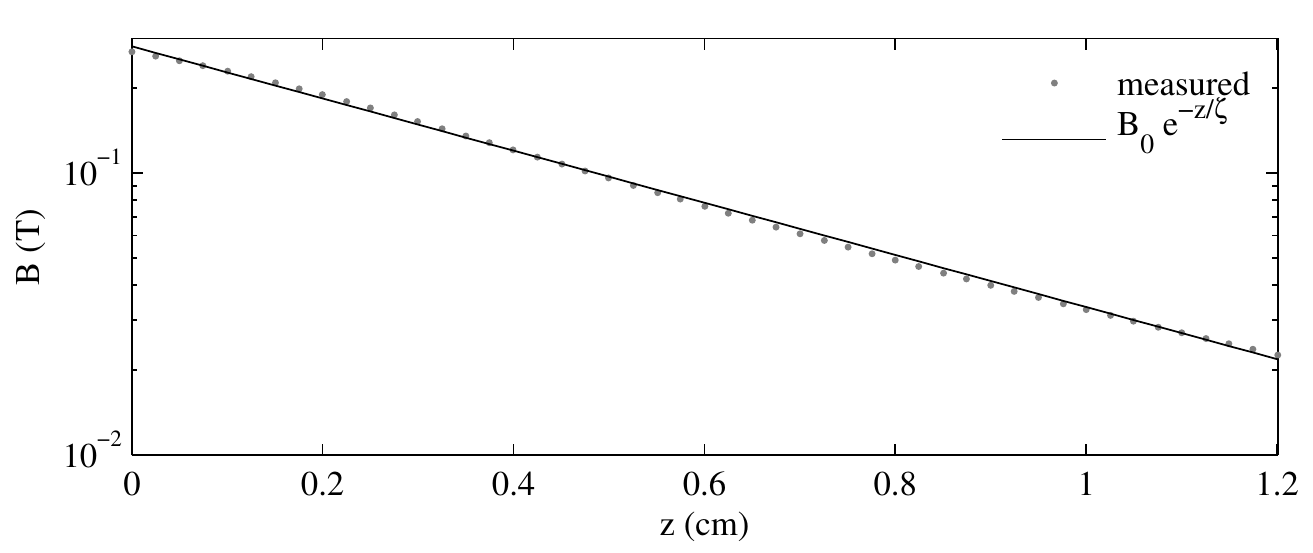}
\caption{\label{fig:Bfalloff} Axial variation of the field of an example magnet, as measured with a Gaussmeter. The plotted exponential fit has $B_0=0.28$~T and $\zeta = 4.70$~mm.}
\end{center}
\end{figure}

Below the floor of the test bed lies a square array of 34 x 34 neodymium-iron-boron (NdFeB) grade N52 magnets, spaced 2.54~cm on center, with alternating polarity. The magnets are cylindrical, with a diameter of 12.7~mm and a thickness of 3.2~mm. Each magnet has a field that decays exponentially with the axial distance from its surface (see figure~\ref{fig:Bfalloff}), with a peak value near 0.3~T and a decay length of $\zeta=4.70$~mm. A pair of bar electrodes are mounted on opposite ends of the test bed and connected to a BK Precision 1670A power supply, allowing an electric current to be passed through the saltwater layer. That current interacts with the magnetic fields via the Lorentz force $\bm{F} = \bm{J} \times \bm{B} / \rho$, where $\bm{J}$ is the current density and $\bm{B}$ is the magnetic field, to produce bulk motion in the salt water. For all experiments discussed below, the forcing current was steady in time. At low current density $J=|\bm{J}|$, the forcing is gentle and the flow generated is a steady lattice of vortices of (spatially) alternating direction. At higher currents, stronger forcing leads to time-dependent flows, loss of lattice symmetry, spatiotemporal chaos, and perhaps two-dimensional turbulence. We characterize the flow by its Reynolds number
\begin{equation}
Re = \frac{u_{rms} L_f}{\nu},
\end{equation}
where $L_f=2.54$~cm is the forcing scale and $u_{rms} = \sqrt{ \left< u^2 + v^2 \right> }$ is the root-mean-square velocity, with $\left< \cdot \right>$ signifying an average over all particles in each frame. Here and throughout we represent the two-dimensional flow as $\bm{u} = u \hat{\bm{x}} + v \hat{\bm{y}}$, adopting Cartesian coordinates $(x,y,z)$ with unit vectors $(\hat{\bm{x}},\hat{\bm{y}},\hat{\bm{z}})$, respectively, $\hat{\bm{x}}$ being the direction of the forcing current and $\hat{\bm{z}}$ being the out-of-plane direction. Flows with $50 \le Re \le 250$ are accessible with our current experimental setup. 

To the saltwater layer we add tracer particles whose motion indicates that of the fluid. The particles are fluorescent polystyrene spheres, 51~$\mu$m in diameter. Their specific gravity is 1.05, intermediate between that of the saltwater layer and the freshwater layer. Thus they are trapped at the interface between the layers. Due to the miscibility of the two fluid layers, there are no long-range surface-tension forces between the particles. In contrast, if the particles are allowed to float to an interface between salt water and air, they quickly clump~\cite{Vella:2005} and become poor tracers, with a much larger effective diameter. The particles absorb most strongly in the blue (468~nm) and emit most strongly in the green (508~nm). We illuminate them with blue light emitting diodes (LEDs) whose luminosity peaks at 470~nm. We subsequently image the particles using an IDT MotionPro M5 camera, first filtering optically to attenuate wavelengths below 520~nm, thereby reducing blue glare much more than the green light emitted by the particles. The camera has a CMOS sensor with pixel dimensions 2320 x 1728 and can record continuously at frame rates up to 170~Hz. We trigger each frame externally, using an Agilent 33210A function generator. The resulting movies are recorded to disk for later processing and particle tracking. 

\subsection{Particle Tracking and Velocimetry}

To measure the flow dynamics, we identify and follow the particles in recorded movies using Lagrangian particle tracking \citep{Ouellette:2006,Ouellette:2010}. As is done in classical PTV, we locate the positions of individual particles and match them in time. Unlike in PTV, however, which typically only uses two images to match particles, we keep the full Lagrangian information about the long-time particle trajectories. Lagrangian dynamics are a powerful way to study fluid flow \citep{Toschi:2009}; for the present application, the length of our trajectories is useful only in that it allows us to use a numerical differentiation scheme that is more robust than simple finite differences. 

Our tracking algorithm has been described in detail elsewhere \citep{Ouellette:2006,Ouellette:2010}, and so here we only briefly summarize the major steps. Particles are located by using local maxima in image intensity (above a small threshold) and the pixels immediately adjacent to the maxima. We obtain the particle centers with a resolution of roughly 0.1 pixels (13.6~$\mu$m in the experiments described herein) by fitting one-dimensional Gaussians to the horizontal and vertical intensity profiles. Once the particle locations have been found, they are tracked using a predictive three-frame best-estimate algorithm. For each partially constructed trajectory, the expected position of the particle at the next time step is estimated using simple kinematics. The measured particle position that comes closest to this estimate is then chosen to continue the trajectory. Small temporal gaps (due to particle drop-out) are bridged via extrapolation. The entire process has been shown to be robust under a wide range of flow conditions, and has been parallelized so as to be computationally efficient. For the data reported here, we tracked $N_p \sim 20~000$ particles per frame. We subsequently differentiate the trajectories temporally by convolving the measured trajectories with a Gaussian smoothing and differentiating kernel \citep{Mordant:2004a,Ouellette:2007a}, yielding a time series of positions and velocities for each tracked particle. 

\section{Velocity Reconstruction}
\label{sec:reconstruction}

After performing Lagrangian particle tracking and differentiating the resulting trajectories, we are left with a collection of particle tracks that, taken together, approximate the velocity field of the fluid. The state of the art in imaging technology and computer processing is such that we can produce data sets whose size and level of detail approaches that of numerical simulation. Other characteristics of our data sets, however, are unlike numerical data~--- in many ways they resemble the observational data of oceanographers. First, our velocity measurements do not lie on a regular grid, but lie at locations that change from frame to frame as the tracer particles follow the flow. The spatial distribution of the tracers also tends to be inhomogeneous. Second, outlier points unavoidably appear in the data. A small fraction are due to physical imperfections in the experimental apparatus, such as dust particles floating on the water surface. More arise through the course of tracking particles and differentiating tracks. Whatever their source, clearly non-physical outlier points require correction. Third, a small but finite compressibility appears in our velocity measurements. Since the typical flow velocities are on the order of 1~cm/s and the speed of sound in water is 1484~m/s, we do not expect physical compressibility; rather, the apparent compressibility is due partly to vertical flow and largely to numerical noise inherent in spatial gradients calculated from irregular samples at finite resolution. These three characteristics, common to any experimental data gathered by PTV or by Lagrangian particle tracking, present challenges and highlight the need for careful post-processing. 

Perhaps there are other ways to sidestep these three challenges. PIV measures fluid velocities on a regular grid, avoiding the first challenge, but PIV is necessarily Eulerian. Interpolating the velocity field onto a regular grid is straightforward but requires smoothing with parameters that are typically empirical and semi-arbitrary. Perhaps worse, interpolation can yield more apparent information than has actually been measured. Adjusting the algorithmic parameters used in identifying and tracking particles can reduce the number of outliers substantially, addressing the second challenge. Outliers cannot be eliminated entirely, however, and reducing their count means an even larger reduction in the count of ``good'' particles being tracked~--- valuable data is lost~\citep{Ouellette:2006,Ouellette:2010}. A variety of experimental techniques have been developed to reduce vertical flow in devices like ours, and we employ many such techniques to address the third challenge. Interpolating and smoothing might lead to spatial gradients with less numerical noise and thus a lower apparent compressibility, but the choice of smoothing and interpolation parameters remains empirical at best. 

The post-processing technique presented below addresses these challenges while producing a final data set that more directly resembles the experimental measurements, relying on few physical or numerical assumptions. Moreover, it allows for multi-scale reconstruction of the velocity field at specified length scales. The remainder of this section details the technique, first outlining the construction of basis modes for the velocity components from a velocity potential function and a streamfunction, then describing numerical techniques for calculating spatial gradients without interpolation. Continuing, we present the projection algorithm, show how to account for open boundaries with velocity potential modes, address the choice of basis for the streamfunction modes, and close with a concise summary of the algorithm. 

\subsection{Velocity Basis}

Two-dimensional, incompressible vector fields can always be represented in terms of a streamfunction, a mathematical fact that finds frequent use in the study of fluids. Its validity arises from the Helmholtz-Hodge theorem: any vector field that vanishes at its boundary (possibly at infinity) can be uniquely decomposed into an incompressible component and an irrotational component~\citep{Arfken:2001}. In two dimensions,
\begin{equation}
\label{eqn:helmholtz}
\bm{u} = \bm{\nabla} \Phi - \hat{\bm{z}} \times \bm{\nabla} \Psi,
\end{equation}
where
\begin{equation}
\bm{\nabla} = \hat{\bm{x}} \frac{\partial}{\partial x} + \hat{\bm{y}} \frac{\partial}{\partial y}.
\end{equation}
In the specific case where $\bm{u}$ is a velocity field, $\Phi$ is known as the velocity potential and $\Psi$ as the streamfunction. The first term on the right-hand side of equation~\ref{eqn:helmholtz} is irrotational in two dimensions, and the second is solenoidal in two dimensions, that is,
\begin{eqnarray}
\bm{\nabla} \times \left( \bm{\nabla} \Phi \right) & = & 0, \nonumber \\
\bm{\nabla} \cdot \left( \hat{\bm{z}} \times \bm{\nabla} \Psi \right) & = & 0. \nonumber
\end{eqnarray}

Alternatively, the streamfunction may be understood in terms of toroidal and poloidal components. Any three-dimensional, incompressible vector field can be decomposed uniquely into a toroidal and a poloidal component, as is commonly done in geophysics, for example by~\citet{Bullard:1954}. The symmetry axis of the toroidal and poloidal components may be chosen \textit{a priori} and should correspond to the symmetry of the system being considered; with in-plane flow in mind, we can choose the out-of-plane vector $\hat{\bm{z}}$ and write the decomposition as~\citep{Chu:2003}
\begin{equation}
\label{eqn:toroidalpoloidal}
\bm{u} = \bm{\nabla}_3 \times \hat{\bm{z}} \Psi + \bm{\nabla}_3 \times \bm{\nabla}_3 \times \hat{\bm{z}} \chi,
\end{equation}
where $\Psi$ and $\chi$ are the toroidal and poloidal scalar functions, respectively, and 
\begin{equation}
\bm{\nabla}_3 = \hat{\bm{x}} \frac{\partial}{\partial x} + \hat{\bm{y}} \frac{\partial}{\partial y} + \hat{\bm{z}} \frac{\partial}{\partial z}.
\end{equation}
In the special case of incompressible, two-dimensional flow, $\chi=0$ and $\Psi$ again becomes the streamfunction. 

These mathematical identities immediately allow for sophisticated manipulation of velocity fields that are already incompressible and two-dimensional. They also allow for conditioning of imperfect, real-world data by representing those measurements in terms of a two-dimensional, incompressible basis~--- because such a basis can be built by decomposing the velocity potential function and streamfunction. Following \citet{Chu:2003} and \citet{Lekien:2004}, we write the velocity potential function as
\begin{equation}
\label{eqn:phidecomp}
\Phi(x,y) = \sum_{j=1}^{N_b} \alpha_j \phi_j(x,y),
\end{equation}
with some coefficients $\alpha_j$, for some number $N_b$ of orthogonal basis functions $\phi_j(x,y)$, which we will call the ``boundary modes'' and usually abbreviate $\phi_j$. Likewise we write the streamfunction as
\begin{equation}
\label{eqn:psidecomp}
\Psi(x,y) = \sum_{k=1}^{N_i} \beta_k \psi_k(x,y),
\end{equation}
with some coefficients $\beta_k$, for some number $N_i$ of orthogonal basis functions $\psi_k(x,y)$, which we will call the ``interior modes'' and usually abbreviate $\psi_k$. Choosing $N_b = N_i = \infty$ allows exact representation of any velocity potential function and any streamfunction, but in numerical calculations it is necessary to truncate the series by choosing $N_b$ and $N_i$ finite. The choice of $N_b$ and $N_i$, as well as the exact forms of the boundary modes $\phi_j$ and the interior modes $\psi_k$, will be specified below.

As mentioned above, one hypothesis of the Helmholtz-Hodge theorem (equation~\ref{eqn:helmholtz}) requires that the amplitude of the vector field be zero at its boundary. However, we typically observe particle velocities in a central sub-region of our experimental apparatus, such that the rigid sidewalls are not in view, and flow through the boundaries of the measurement region is not only possible but ever-present. Thus the boundary requirement is not satisfied. This situation is common in experimental and observational studies, where measurements are often recorded in a region smaller than the fluid container or oceanic coastline. Mathematically, such boundaries are said to be ``open.'' With open boundaries, decomposition into incompressible and irrotational components remains possible, but is no longer unique~\citep{Lynch:1989}. Rather, the results of decomposition depend on the boundary conditions, which can be chosen freely. Following~\citet{Lekien:2004}, we choose the boundary condition
\begin{equation}
\label{eqn:bdrycond}
\left. \Psi \right|_\Gamma=0,
\end{equation}
implying also $\left. \psi_k \right|_\Gamma=0$, where $\Gamma$ represents the boundary. In words, all inflow and outflow at the boundary is accounted by the velocity potential function $\Phi$, not the streamfunction $\Psi$. 

With equations~\ref{eqn:helmholtz} and \ref{eqn:bdrycond}, the Helmholtz-Hodge decomposition of the velocity is specified uniquely, and with equations~\ref{eqn:psidecomp} and \ref{eqn:phidecomp}, the streamfunction and velocity potential function have been constructed. However, in experiments we do not measure the streamfunction or velocity potential function, so we cannot project our data onto them directly. Rather we build basis sets for the two horizontal components of the velocity by using equation~\ref{eqn:helmholtz} with each basis mode, yielding
\begin{eqnarray}
\label{eqn:velocitybases}
u & = & \sum_{j=1}^{N_b} \alpha_j \frac{\partial \phi_j}{\partial x} + 
\sum_{k=1}^{N_i} \beta_k \frac{\partial \psi_k}{\partial y}, \\
v & = & \sum_{j=1}^{N_b} \alpha_j \frac{\partial \phi_j}{\partial y} -
\sum_{k=1}^{N_i} \beta_k \frac{\partial \psi_k}{\partial x}. \nonumber
\end{eqnarray}
Hence we have constructed incompressible, two-dimensional basis sets onto which measurements can be projected. Since the velocity components $u$ and $v$ are constructed from the same velocity potential function $\Phi$ and same streamfunction $\Psi$, the coefficients $\alpha_j$ and $\beta_k$ are the same for both; see equation~\ref{eqn:leastsquares}.

\subsection{Gradients at Particle Locations}

As equation~\ref{eqn:velocitybases} makes explicit, constructing the velocity basis from the velocity potential function basis $\phi_j$ and the streamfunction basis $\psi_k$ requires calculating their spatial gradients. If $\phi_j$ and $\psi_k$ were known analytically, their gradients could be calculated exactly. On the other hand, if $\phi_j$ and $\psi_k$ are constructed numerically, their gradients must be approximated. As emphasized above, our intention is to condition experimental measurements and gain access to their properties at various length scales, while maintaining as much fidelity to the original measurements as possible. We would like to minimize the number of post-processing parameters introduced, avoiding interpolation and smoothing. Our method for numerical calculation of gradients must follow suit. 

\begin{figure}
\begin{center}
\includegraphics[width=\textwidth]{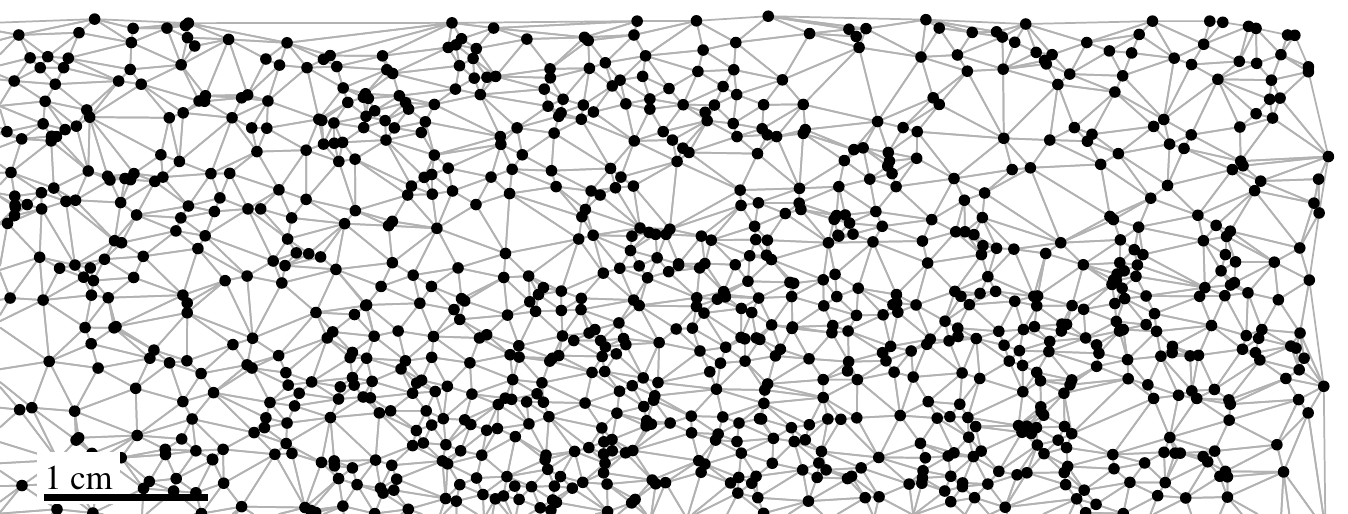} 
\caption{\label{fig:particlemesh} A portion of an example particle mesh as seen from above. Particle locations are indicated by black dots much larger than the actual size of the particles. The mesh itself, computed from the particle locations via a Delaunay triangulation, is shown in grey.}
\end{center}
\end{figure}

Our observations, however, are not made at spatial locations that are regular or predictable~--- we measure the fluid velocity at the particle locations (which are effectively random) and nowhere else, as figure~\ref{fig:particlemesh} shows. Thus grid-based methods for calculating gradients, such as simple finite differences, would require interpolation onto a regular grid and are inappropriate here. When it becomes necessary to calculate numerical gradients for use in equation~\ref{eqn:velocitybases} or in equations below, we do so \textit{at the particle locations} via a finite-element technique. Typically, finite-element methods are used to solve partial differential equations on the vertices of a triangular ``mesh'' that is constructed \textit{a priori} and often refined iteratively to produce sufficiently low numerical error. Our approach differs: the mesh is not constructed nor refined \textit{a priori}, but is specified by the particles themselves, whose locations are the vertices of the mesh. Its edges are constructed via a Delaunay triangulation, after which we remove edge triangles whose aspect ratios are too small (below 0.1) and who are therefore prone to numerical instability. Consequently, the velocity bases given by equation~\ref{eqn:velocitybases}, like all other quantities, are known at the particle locations, and neither interpolation nor smoothing is necessary. A new mesh like the one shown in figure~\ref{fig:particlemesh} is constructed for each frame of data, since particles move between frames. 

\subsection{Least-Squares Projection}

Once incompressible, two-dimensional bases have been constructed at the particle locations for both components of the velocity, the next task is to determine the coefficients $\alpha_j$ and $\beta_k$ in equation~\ref{eqn:velocitybases}. They can be calculated by projecting measurements from tracked particles onto the bases using a linear least-squares method. For notational simplicity, we define the full set of basis functions (including contributions from both the velocity potential function and the streamfunction) as
\begin{eqnarray}
\label{eqn:fullbases}
u_l & = & \left\{ \frac{\partial \phi_j}{\partial x} \right\} \cup 
\left\{ \frac{\partial \psi_k}{\partial y} \right\}, \\
v_l & = & \left\{ \frac{\partial \phi_j}{\partial y} \right\} \cup 
\left\{ - \frac{\partial \psi_k}{\partial x} \right\}, \nonumber
\end{eqnarray}
and the full set of coefficients as $\gamma_l = \left\{ \alpha_j \right\} \cup \left\{ \beta_k \right\}$. Then equation~\ref{eqn:velocitybases} can be rewritten as
\begin{eqnarray}
\label{eqn:velocitybases2}
u & = & \sum_{l=1}^{N_b+N_i} \gamma_l u_l, \\
v & = & \sum_{l=1}^{N_b+N_i} \gamma_l v_l. \nonumber
\end{eqnarray}
The total squared error of the projection is then the square of the difference between equation~\ref{eqn:velocitybases2}, evaluated at the particle locations, and the measurements themselves. Minimizing that error is mathematically identical to maximizing the probability that the coefficients $\gamma_l$ (or equivalently, $\alpha_j$ and $\beta_k$) correctly model the measured data, assuming Gaussian errors. It can be shown~\citep{Press:2007} that the error is minimized by the coefficients $\gamma_l$ that satisfy the matrix equations
\begin{equation}
\label{eqn:leastsquares}
\sum_{p=1}^{N_p} \sum_{l=1}^{N_b+N_i} \left( u_{lp} u_{jp} + v_{lp} v_{jp} \right) \gamma_l = \sum_{p=1}^{N_p} \left( u_{jp} u_p^{meas} + v_{jp} v_p^{meas} \right),
\end{equation}
known as the ``normal equations'' of the linear, least-squares fit. Here $p$ indexes the $N_p$ particles present in the frame, and all sums are written explicitly. The notation $u_{lp}$ signifies the value of the basis mode $u_l$, as defined in equation~\ref{eqn:fullbases}, evaluated at the location of particle $p$; $u_{jp}$, $v_{lp}$, and $v_{jp}$ have corresponding meanings. Likewise $u_p^{meas}$ and $v_p^{meas}$ are the components of the velocity as measured at particle $p$. 

To solve equation~\ref{eqn:leastsquares}, we use singular value decomposition, a mathematical operation that avoids numerical issues that arise due to nearly singular matrices and due to basis modes that are nearly degenerate at the particular locations where measurements have been recorded~\citep{Press:2007}. Once the coefficients $\gamma_l$ are known, the velocity components $u$ and $v$ can be reconstructed by simply summing over the basis modes, that is, by using equation~\ref{eqn:velocitybases2} directly.

\subsection{Boundary Modes}

So far we have required only that the basis functions $\phi_j$ and $\psi_k$ be orthogonal, without specifying them exactly; we now return to the question of choosing those bases, first discussing the boundary modes $\phi_j$, then the interior modes $\psi_k$. 

To construct a basis for the velocity potential function, we follow the method of~\citet{Lekien:2004}. In an incompressible, two-dimensional flow with closed boundaries, $\Phi=0$ identically. We retain the velocity potential function, however, to account for inflow and outflow through the open measurement boundaries~--- a choice that led to equation~\ref{eqn:bdrycond}. Thus the boundary modes are determined completely by the boundary conditions. We denote the local outflow at the boundary $\Gamma$ with the function $G(\tau) = \hat{\bm{n}} \cdot \bm{u}$, where $\hat{\bm{n}}$ is the unit vector locally normal to the boundary and pointing outward, and the coordinate $\tau$ gives the distance around the boundary, measured from some arbitrary starting point. The local outflow can be decomposed in Fourier modes as 
\begin{equation}
\label{eqn:outflow}
G(\tau) = \sum_{j=1}^{N_b} \mu_j g_j \left( \frac{\pi \tau}{L_b} \right),
\end{equation}
with some coefficients $\mu_j$, for some number $N_b$ of modes, where $L_b$ is the total length of the boundary (so $0 \le \tau < L_b$) and 
\begin{equation}
\label{eqn:outflowbasis}
g_j(x) = \left\{
\begin{array}{ll}
\sin{jx}, & j~\mbox{even}, \\
\cos{(j+1)x}, & j~\mbox{odd}.
\end{array} \right.
\end{equation}
We choose the basis this way so that each $g_j$ is continuous over the boundary and 
\begin{equation}
\int_0^{L_b} G(\tau) d\tau = \int_0^{L_b} g_j \left( \frac{\pi \tau}{L_b} \right) d\tau = 0,
\end{equation}
as required by incompressibility. 

\begin{figure}
\begin{center}
\includegraphics[width=\textwidth]{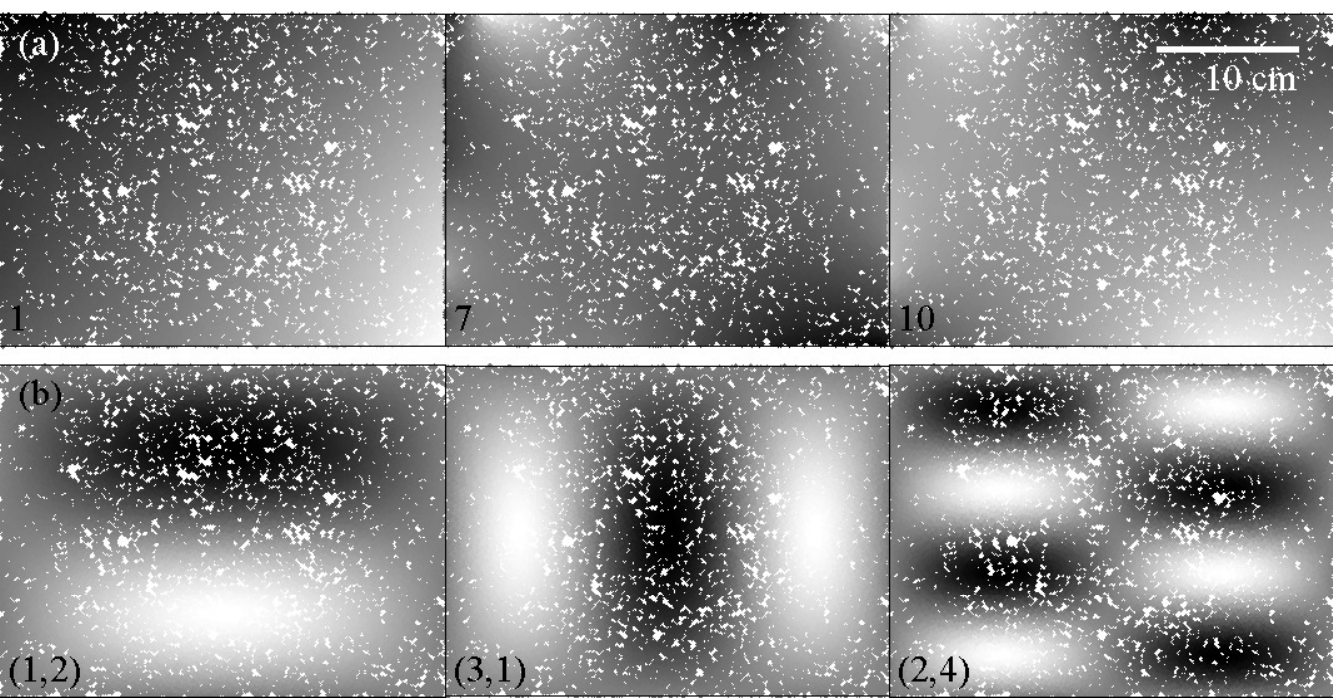} 
\caption{\label{fig:bases} Example basis modes. (a) Three boundary modes $\phi_j$. A dot is drawn at the location of each tracked particle, much larger than the actual particle, with its shade representing the value of the mode at that location. Dark shades signify positive values and light shades, negative. The amplitude is arbitrary, and the white background shows through in regions where few particles are present. The particles are seen from above, and the forcing current flows from left to right. The value of $j$ is indicated in each case. (b) Three interior modes $\psi_k$, drawn in the same way. The values of $(n_x,n_y)$ are indicated in each case; see equation~\ref{eqn:fourierbasis}. The velocity bases $(u_l,v_l)$ are constructed from spatial gradients of $\phi_j$ and $\psi_k$ according to equation~\ref{eqn:fullbases}.}
\end{center}
\end{figure}

We construct the basis modes for the velocity potential function using solutions of Laplace's equation, with the outflow modes $g_j(\pi \tau / L_b)$ providing the Neumann boundary condition:
\begin{eqnarray}
\label{eqn:phibasis}
\nabla^2 \phi_j & = & 0 \\
\left. \hat{\bm{n}} \cdot \bm{\nabla} \phi_j \right|_\Gamma & = & 
\left. g_j \left( \frac{\pi \tau}{L_b} \right) \right|_\Gamma . \nonumber
\end{eqnarray}
That the modes $\phi_j$ must satisfy Laplace's equation is also a requirement of incompressibility, made clear by taking the divergence of both sides of equation~\ref{eqn:helmholtz}. Three example basis boundary modes, constructed this way, are shown in figure~\ref{fig:bases}. To set $N_b$, the total number of boundary modes, we choose a minimum length scale $L_{min}$ and retain all modes whose scales $L_b$ satisfy
\begin{equation}
\label{eqn:lminboundary}
\frac{2 L_b}{j} \ge L_{min}.
\end{equation}

\subsection{Interior Modes}

To construct a basis for the streamfunction, we use the eigenfunctions of the Laplacian, satisfying 
\begin{equation}
\label{eqn:eigval}
\nabla^2 \psi_k = \lambda_k \psi_k,
\end{equation}
where $\lambda_k$ is the scalar eigenvalue of the eigenfunction $\psi_k$. The eigenfunctions of the Laplacian are used in many situations where an orthogonal, complete set is required; their exact form varies with the geometry and conditions at the boundary. Some examples include spherical harmonics, vector spherical harmonics (toroidal and poloidal components, as in equation~\ref{eqn:toroidalpoloidal}), and Bessel functions. For a scalar function (such as the streamfunction $\Psi(x,y)$) in a two-dimensional, rectangular region, the eigenfunctions of the Laplacian are two-dimensional Fourier modes (products of sines and cosines). 

For geometries in which eigenfunctions of the Laplacian are not known analytically, they may be calculated numerically. Such an approach allows analysis of data sets with boundaries of arbitrary geometry, making it valuable for the irregular observation regions common in oceanography. Its drawback, however, is substantial computational expense. In reconstructing observed velocity fields, it is necessary to choose between analytical and numerical computation of eigenfunctions (or ``eigenmodes''), weighing the strengths and weaknesses of each. Below we compare and contrast reconstructions of the same data set using both techniques. 

The two-dimensional Fourier basis for the streamfunction takes the form
\begin{equation}
\label{eqn:fourierbasis}
\psi_k = \sin \left( \frac{n_x \pi x}{L_x} \right) \sin \left( \frac{n_y \pi y}{L_y} \right),
\end{equation}
where $L_x$ and $L_y$ are the sizes of the rectangular measurement region in the $x$ and $y$ directions, respectively, and $k$ stands for the combination of $n_x$ and $n_y$, both of which are positive integers. Three such modes are shown in figure~\ref{fig:bases}. Note that the eigenvalue equation (\ref{eqn:eigval}) and the boundary condition (equation~\ref{eqn:bdrycond}) are both satisfied automatically; cosine modes, on the other hand, would not satisfy the boundary condition. Fourier modes have well-defined length scales in both dimensions, growing smaller as the integers $n_x$ and $n_y$ increase. To set $N_i$, the total number of interior modes, we employ a criterion similar to \ref{eqn:lminboundary}, retaining all modes whose scales are larger than the minimum in both dimensions:
\begin{eqnarray}
\label{eqn:lminfourier}
\frac{2 L_x}{n_x} & \ge & L_{min}, \\
\frac{2 L_y}{n_y} & \ge & L_{min}. \nonumber
\end{eqnarray}

\begin{figure}
\begin{center}
\includegraphics[width=\textwidth]{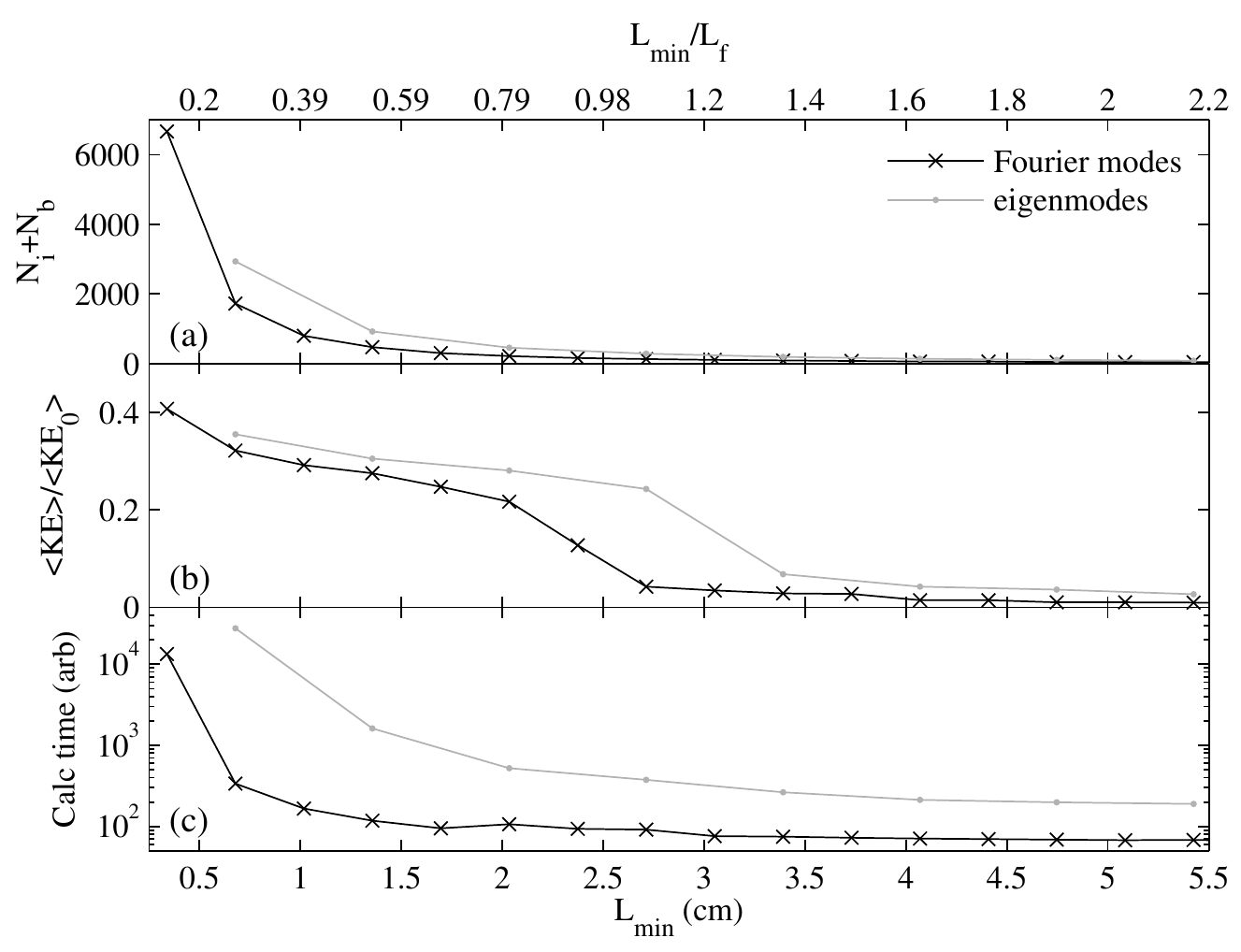} 
\caption{\label{fig:manyLmin} Characteristics of example bases built with varying values of the minimum feature size $L_{min}$, including (a) the total number of basis modes ($N_i$ interior modes and $N_b$ boundary modes), (b) the fraction of kinetic energy captured by the reconstruction, and (c) the time required to build the bases and calculate the reconstruction for one frame, composed of 19~000 particles. Here $\left< KE \right>$ is the mean particle kinetic energy after reconstruction, $\left< KE_0 \right>$ is the mean particle kinetic energy as measured (before reconstruction), and $Re=100$.}
\end{center}
\end{figure}

With the criteria~\ref{eqn:lminfourier} and \ref{eqn:lminboundary}, we have all the information necessary to use the boundary basis obtained from equation~\ref{eqn:phibasis} and the interior basis defined in equation~\ref{eqn:fourierbasis} to reconstruct a velocity field. The effects of varying $L_{min}$ are summarized in figure~\ref{fig:manyLmin}. Smaller values of $L_{min}$ increase $N_i$ and $N_b$, capturing more of the information recorded in the original measurements (which we roughly estimate using kinetic energy), but also requiring longer computation times and additional storage. One noteworthy feature is the large jump in kinetic energy content near the forcing scale. Consistent with intuition, this jump suggests that in order to capture important features in the flow, one should choose $L_{min} < L_f$. Calculation time is plotted on a semi-logarithmic scale~--- its dependence on $L_{min}$ is steeper than exponential. The calculation time also depends strongly on the particle count $N_p$ (not plotted). 

Alternatively, a basis for the interior modes can be constructed by solving equation~\ref{eqn:eigval} numerically. We do so using a finite-element technique similar to the one described above in order to obtain numerical basis functions at the particle locations, without interpolation or smoothing. As in the case of Fourier modes, it is necessary to decide how many modes $N_i$ are required, which can be done by estimating the length scale of each numerical eigenmode using dimensional arguments~\citep{Lekien:2004}. We set $N_i$ by retaining all modes with 
\begin{equation}
\label{eqn:lmineig}
L_1 \sqrt{ \frac{\lambda_1}{\lambda_k}} \ge L_{min},
\end{equation}
where $\lambda_1$ is the lowest eigenvalue and $L_1$ is the boundary length scale (we use $L_1 = L_x$). With the criteria~\ref{eqn:lminboundary} and \ref{eqn:lmineig} we can use the boundary basis obtained from equation~\ref{eqn:phibasis} and the interior basis obtained from equation~\ref{eqn:eigval} to reconstruct the velocity field. The choice of $L_{min}$ has similar effects as with the Fourier basis described above, as shown in figure~\ref{fig:manyLmin}. The kinetic energy fraction again increases sharply near the forcing scale, as one might expect. 

We are now in a position to compare and contrast Fourier modes and numerical eigenmodes as the choice for the interior modes. As stated above, in a rectangular measurement region, the Fourier modes satisfy equation~\ref{eqn:eigval} exactly, and we would expect no difference between the two bases. But even though our camera has a rectangular field of view, our measurement region is not rectangular~--- rather, its edges are set by the locations of the outermost particles, as shown in figure~\ref{fig:particlemesh}. Thus the Fourier basis and the eigenmode basis differ. Figure~\ref{fig:manyLmin} shows that the value of $L_{min}$ where kinetic energy suddenly increases is slightly higher for eigenmodes than for Fourier modes, suggesting that eigenmodes reproduce the flow more efficiently in terms of $L_{min}$. In terms of calculation time, however, they are much less efficient. Even though more Fourier modes are required to reconstruct a velocity field with sufficient energy content, they require much less computation time (by a factor of five in this flow with 20~000 particles, if we wish to capture 30\% of the measured kinetic energy). Thus for the nearly-rectangular boundaries common in our experiments, Fourier modes form a more attractive basis than eigenmodes. In measurement regions whose boundaries are more irregular, the fidelity of the Fourier basis decreases, making the eigenmode basis more attractive. In a circular measurement region, Bessel modes would be the most attractive basis set, though we do not consider them here. 

\begin{figure}
\begin{center}
\includegraphics[width=\textwidth]{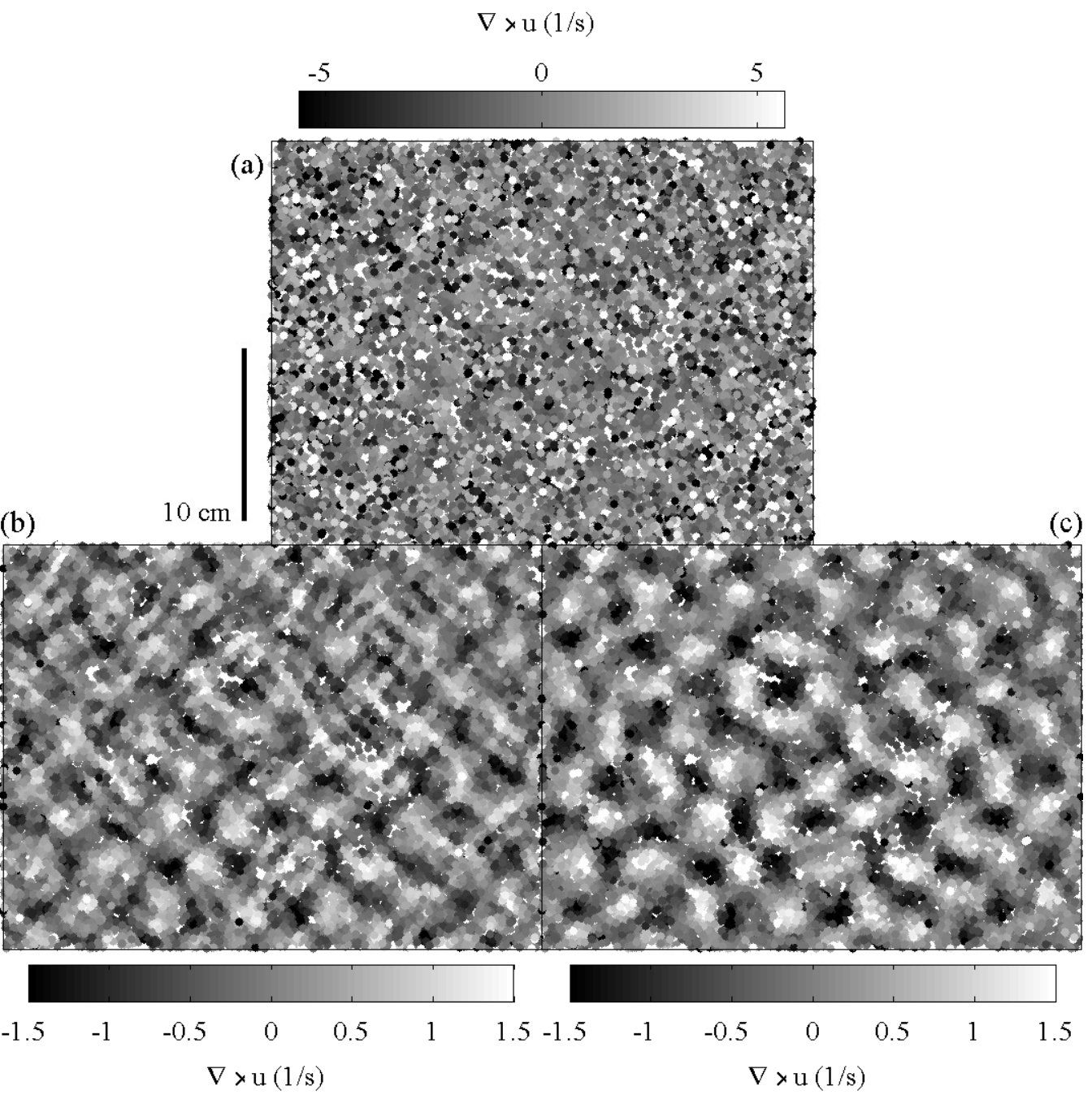} 
\caption{\label{fig:vortcompare} An example vorticity field seen from above (a) as measured, (b) after reconstruction using boundary modes and interior Fourier modes, and (c) after reconstruction using boundary modes and interior eigenmodes. A dot is drawn at the location of each tracked particle, much larger than the actual particle, with its shade representing the value of the vorticity at that location. The white background shows through in regions where few particles are present. For both projections, $L_{min} = 0.27 L_f$. Here $Re=100$, and the forcing current flows from left to right.}
\end{center}
\end{figure}

Another comparison between the Fourier basis and the eigenmode basis is provided in figure~\ref{fig:vortcompare}, which shows the vorticity $\bm{\omega} = \bm{\nabla} \times \bm{u}$ of a single frame as measured and as reconstructed with each basis. By eye, both reconstructions are similar, with noticeable noise reduction when compared to the measured data, though Fourier modes are computationally less expensive. Henceforth we shall use Fourier modes with $L_{min} = L_f/4$. 

The kinetic energy fraction plotted in figure~\ref{fig:manyLmin} does not reach 100\% and perhaps deserves comment. Where is the kinetic energy going, and would interpolation preserve more of it? Particle tracking, like any measurement technique, involves systematic uncertainty which in this case tends to increase the kinetic energy artificially. \citet{Ouellette:2007a} have estimated the uncertainty associated with tracking and differentiating the tracks as
\begin{equation}
\left< \xi_x \left( T \right)^2 \right> = \frac{2.11 \left< \epsilon_x^2 \right>}{T^3},
\end{equation}
where $\left< \xi_x \left( T \right)^2 \right>$ is the velocity variance due to uncertainty, $2T$ is the integration time used when convolving the measured trajectories with the smoothing and differentiating kernel (see \S\ref{sec:experiment}), and $\left< \epsilon_x^2 \right>$ is the variance of the uncertainty in locating particles spatially. In our measurements, $T=1.5$~frames and $\left<\epsilon_x \right>$ is about 0.1~pixels, yielding $\left< \xi_x \left( T \right)^2 \right> = 0.0063$~pixels$^2$/frame$^2$, which corresponds to a standard deviation about 3.3\% of the root-mean-square velocity at $Re=100$. Additional uncertainty arises because of the presence of physical imperfections in the experimental apparatus, such as dust or tracer particles floating on the water surface. Ideally, both types of uncertainty would be eliminated in post-processing. Hence we do not expect the kinetic energy fraction of any reconstruction to reach 100\%. 

\begin{figure}
\begin{center}
\includegraphics[width=\textwidth]{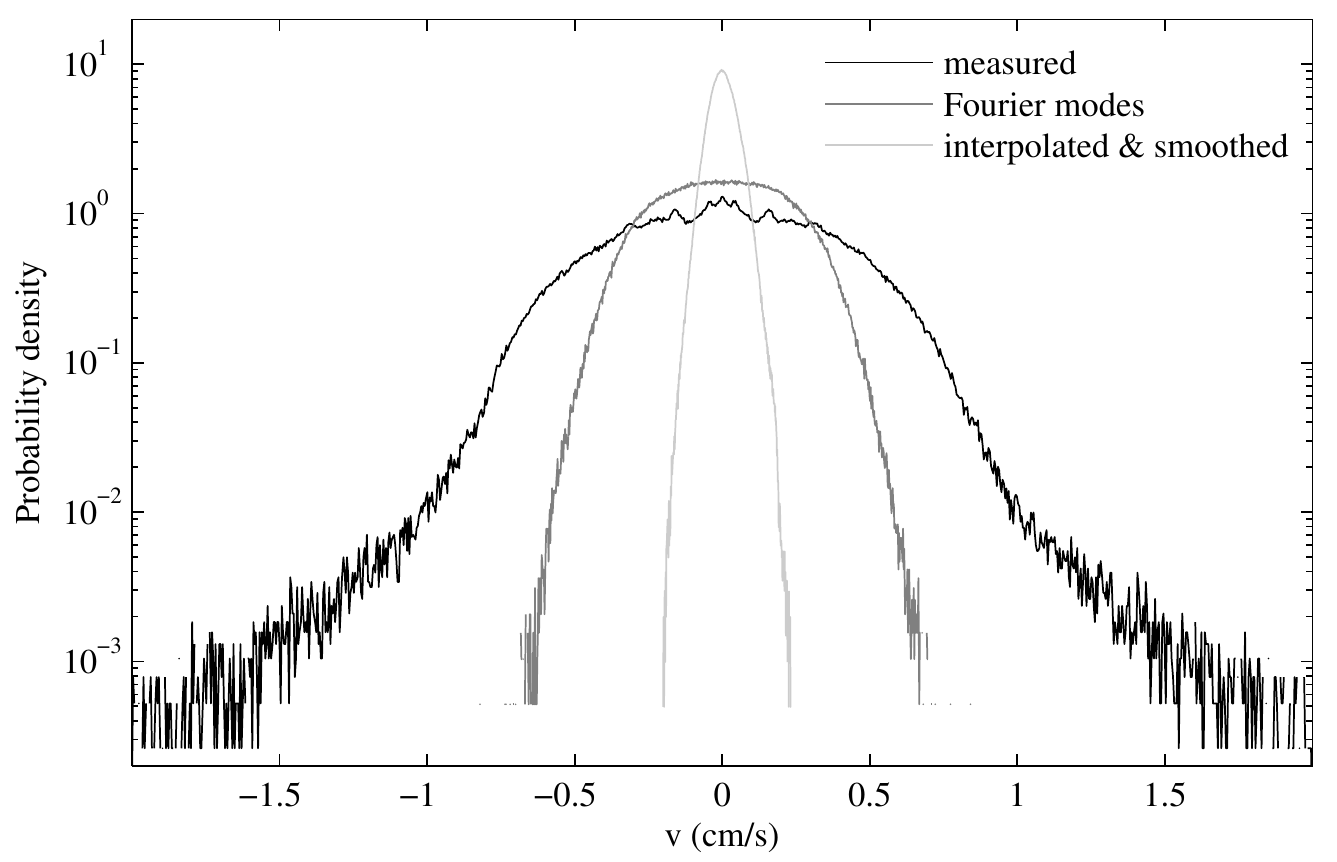} 
\caption{\label{fig:vpdf} Probability density functions of the velocity component $v=\bm{u} \cdot \hat{\bm{y}}$, comparing reconstruction by projection (using the Fourier basis for internal modes) to interpolation and smoothing. In this example projection preserves 33\% of the original kinetic energy, while interpolation and smoothing preserves 2.3\%. The measurement encompasses 50~frames, including about $10^6$ particles, at $Re=100$.}
\end{center}
\end{figure}

One way to examine the effect of reconstruction in more detail is by plotting probability density functions (PDFs) of velocity. We have reconstructed 50~frames of measured data ($Re=100$) by projection with a Fourier basis for the interior modes, using $L_{min}=L_f/4$. We have also reconstructed the same data with interpolation and smoothing, using a smoothing length of $L_f/4$. Figure~\ref{fig:vpdf} shows PDFs of the measured data and both reconstructions. The measured data shows non-Gaussian tails in its velocity PDF which are removed by either reconstruction technique, as one would hope. The kinetic energy fraction is related to the variance of the velocity, indicated by the width of each PDF. As the figure shows, reconstruction by projection results in a velocity variance much closer to the original than does the more common interpolation and smoothing~--- thus more of the kinetic energy is preserved. In this example, projection preserves about 33\% of the kinetic energy, while interpolation and smoothing preserves only about 2.3\%. Thus the kinetic energy fraction plotted in figure~\ref{fig:manyLmin} would be considerably lower if we used interpolation instead of projection. 

\subsection{Summary of the Algorithm}

Readers who are interested in reconstructing other velocity fields using this method may find a concise summary useful. We begin by building a particle mesh via Delaunay triangulation, identifying the edge particles. Next we construct the outflow modes $g_j(\pi \tau / L_b)$ (equation~\ref{eqn:outflowbasis}), using equation~\ref{eqn:lminboundary} to determine the mode count. The $g_j$ give the boundary conditions for a numerical solution of Laplace's equation (equation~\ref{eqn:phibasis}), for which we use finite-element techniques, yielding the boundary modes $\phi_j$. We calculate their gradients numerically to obtain the components of the velocity bases (equation~\ref{eqn:velocitybases}) that arise from the boundary modes. For the special case of closed boundaries (without inflow or outflow), the boundary modes may be ignored altogether, that is, $\phi_j=0$.

The interior Fourier modes $\psi_k$, meanwhile, are given by equation~\ref{eqn:fourierbasis}, and their gradients can be calculated analytically. Thus the components of the velocity bases (equation~\ref{eqn:velocitybases}) that arise from interior modes can be constructed explicitly from trigonometric functions, the mode count being set by equation~\ref{eqn:lminfourier}. With the velocity basis now complete, we solve the normal equations of the linear, least-squares fit (equation~\ref{eqn:leastsquares}) numerically using singular value decomposition to obtain the coefficients $\gamma_l$. Finally we reconstruct the velocity field by summing its components according to equation~\ref{eqn:velocitybases2}. 

\section{Multi-scale Structure}
\label{sec:structure}

\subsection{Spectral Power}

\begin{figure}
\begin{center}
\includegraphics[width=\textwidth]{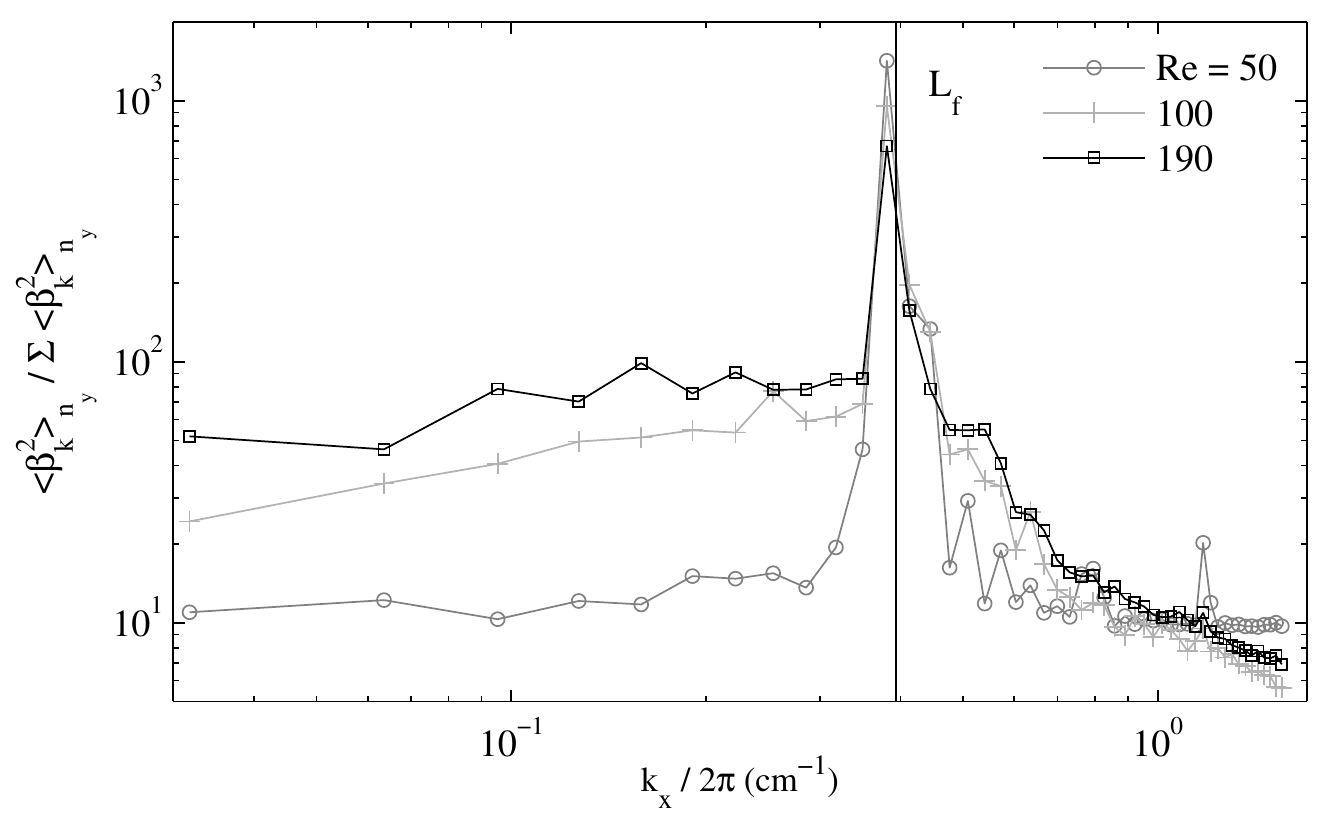} 
\caption{\label{fig:Fourierspec} Normalized power spectral density of the velocity field at various Reynolds numbers, calculated using the coefficients $\beta_k$ of basis modes with well-defined length scales. The forcing scale $L_f$ is indicated with a vertical line. Here $\left< \cdot \right>_{n_y}$ signifies averaging in the $y$ direction, that is, averaging all modes with the same $k_x$.}
\end{center}
\end{figure}

Reconstructing observed velocity fields using the technique outlined in \S\ref{sec:reconstruction} can be useful to condition a data set, removing outlier particles and reducing the apparent compressibility associated with flow in the vertical direction. Because it involves projection onto a basis of orthogonal modes with well-defined length scales, it also offers another benefit, direct access to the multi-scale structure of the flow. For example, instantaneous spatial power spectra can be calculated without interpolation or smoothing~--- the spectral power in each mode is simply the square of its coefficient $\gamma_l$. Producing a one-dimensional spectrum requires averaging over one spatial dimension, and a variety of choices of its direction are possible, including $x$, $y$, and the azimuthal direction. Averaging the spectral power in the $y$ direction and sorting the averaged modes according to their $x$ wavenumber $k_x = 2 \pi n_x / L_x$ (see equation~\ref{eqn:fourierbasis}) yields spectra like those shown in figure~\ref{fig:Fourierspec}. Each is the product of averaging 36 $y$-direction modes in each of 997 frames of data, spanning about 1~min in time and including roughly $10^7$ particles. No interpolation, smoothing, or windowing is required. 

Consistent with our expectations, figure~\ref{fig:Fourierspec} shows a peak in power at the wavenumber nearest the forcing scale $L_f$. The spectra are robust in that they account for a large number of measurements, but their range of scales is insufficient to comment on scaling laws or on the presence or absence of the enstrophy and energy cascades predicted by~\citet{Kraichnan:1967}. Choosing a smaller value for $L_{min}$ would add modes at smaller scales and increase the range, but the increase in computational expense is steep (see figure~\ref{fig:manyLmin}). When a spectrum is desired over a wider range of scales, a better method would be to use criteria different from equation~\ref{eqn:lminfourier} for choosing the number of basis functions in the $x$ and $y$ directions, such as
\begin{eqnarray}
\frac{L_x}{\pi n_x} & \ge & \xi L_{min}, \nonumber \\
\frac{L_y}{\pi n_y} & \ge & L_{min}. \nonumber
\end{eqnarray}
With the dimensionless parameter $\xi < 1$, this basis would include more scales in the $x$ direction without adding more modes in the $y$ direction. Thus the computational expense would remain manageable, and additional variation in $y$ would have been removed in averaging in any case. Since $\xi$ and $L_{min}$ can be adjusted, no limits to the spectral range need be introduced in post-processing; only the limit imposed by the data itself~--- the particle spacing~--- remains. 

\subsection{Scale-Local Velocity}

Spectra give some measure of the multi-scale structure of a flow and are common in the literature because of the relative ease with which they can be obtained, via a variety of methods. Reconstruction by projection onto basis modes gives access to much more information, however, including scale-local velocity fields. Inspired by techniques used in large-eddy simulation, \citet{Rivera:2003} developed a method for calculating the transfer rates of enstrophy and energy directly, using band-pass spatial filters. Later \citet{Bruneau:2007} introduced a similar technique based on wavelets. We suggest that projecting observational data onto basis modes with well-defined length scales can allow for the same sort of direct calculation of enstrophy and energy flux, without the complication of filter construction and tuning. 

\begin{figure}
\begin{center}
\includegraphics[width=\textwidth]{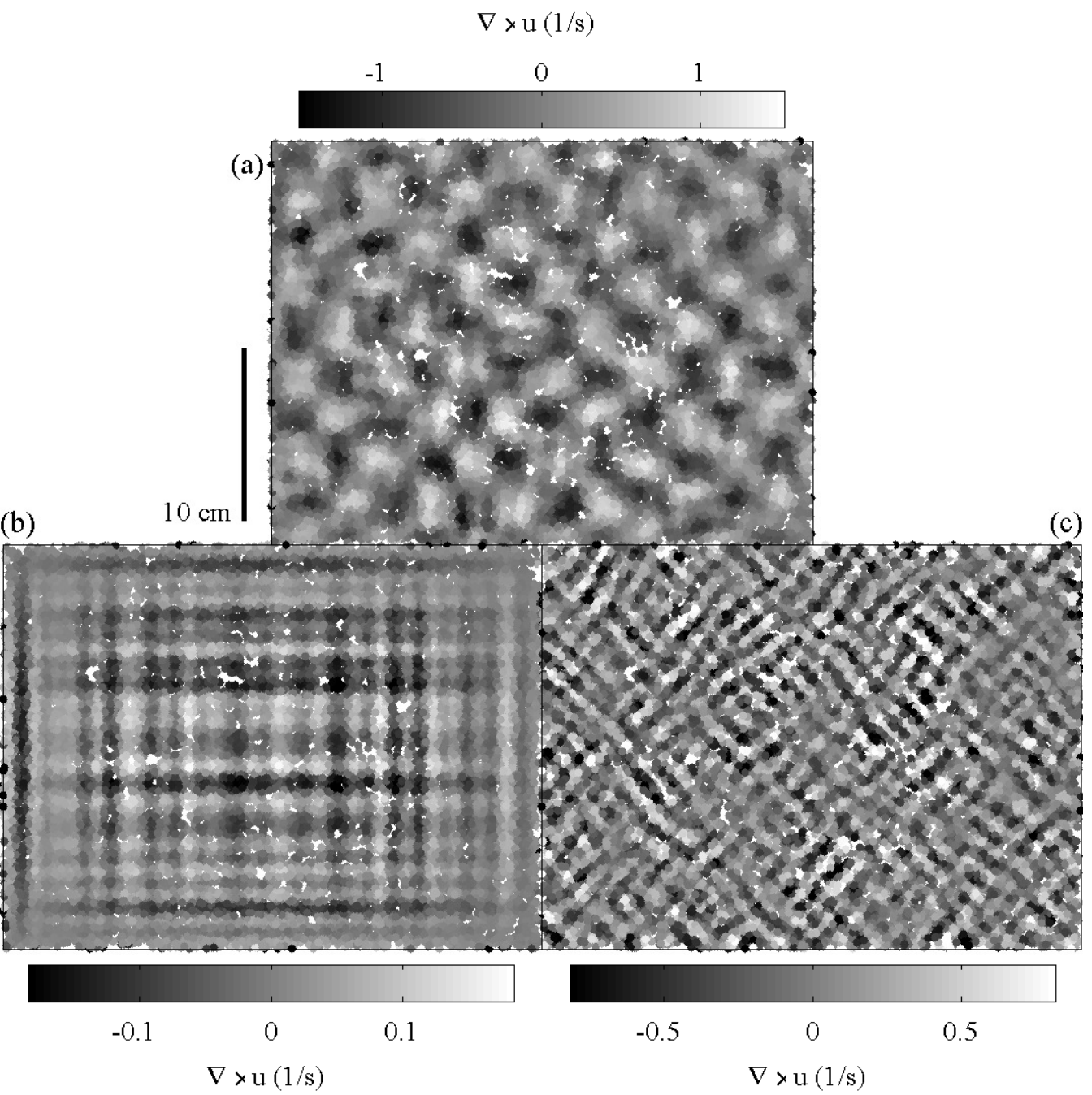} 
\caption{\label{fig:vortscales} Components of the vorticity field shown in figure~\ref{fig:vortcompare}, drawn the same way, at length scales (a) near the forcing scale, with $2 L_f /3 \le L \le 5 L_f$, (b) larger than the forcing scale, with $5 L_f \le L$, and (c) smaller than the forcing scale, with $L \le 2 L_f /3$. Note that the limits of the color scale are different in each image. Here we use a Fourier basis for the interior modes, and the length scale $L$ of each mode is defined according to equation~\ref{eqn:lengthscale}.}
\end{center}
\end{figure}

As a proof of principle, we offer figure~\ref{fig:vortscales}. It reproduces an example vorticity field (the same data as in figure~\ref{fig:vortcompare}), splitting the flow into components with length scales near the forcing scale, larger than forcing scale, and smaller than the forcing scale. Instead of applying band-pass filters, we can select only the part of the velocity field relevant to a particular band of length scales by retaining only those Fourier modes whose length scales
\begin{equation}
\label{eqn:lengthscale}
L = \frac{1}{\sqrt{2}} \sqrt{ \left( \frac{L_x}{n_x} \right)^2 + \left( \frac{L_y}{n_y} \right)^2 }
\end{equation}
fall within the band. The factor $1/\sqrt{2}$ assures that modes with
\begin{equation}
\frac{L_x}{n_x} = \frac{L_y}{n_y} = L_{f}
\end{equation}
have length scale $L = L_f$. Choosing $2 L_f /3 \le L \le 5 L_f$ as the limits of the band near the forcing scale, we find that flow in this range of scales bears a strong resemblance to the full flow as shown in figure~\ref{fig:vortcompare}. One might expect the resemblance not only because we have included the forcing scale but also because the majority of the energy is captured by this band, as indicated by figure~\ref{fig:Fourierspec}. Plotting the vorticity at different length scales shows that the majority of the enstrophy is also captured by the band near $L_f$, as the color scales of figure~\ref{fig:vortscales} make clear. The flow features with scales $L \le 2 L_f / 3$ appear to be dominated by frequency harmonics of the forcing lattice, though other phenomena are present as well. The flow features with scales $L \ge 5 L_f$ are dominated by structures that are large in one spatial dimension but small in the other. The power associated with circulations that are large in both spatial dimensions is small compared to other flow components. 

\subsection{Boundary Effects}

In addition to giving access to scale-local spectral power and velocity (and vorticity) components, the reconstruction method described in \S\ref{sec:reconstruction} allows separation of the scales and structures associated with boundary effects from those associated with interior flow. For two-dimensional, incompressible flows without outflow at the boundaries, the streamfunction as decomposed in equation~\ref{eqn:psidecomp} with the boundary condition~\ref{eqn:bdrycond} comprises a sufficient representation, and we need no more components. In the case of a flow with open boundaries, the velocity potential modes (equation~\ref{eqn:phidecomp}) are included as a necessary and sufficient account of the outflow. Since the flows associated with velocity potential modes $\phi_j$ are orthogonal to those associated with streamfunction modes $\psi_k$, their motions may be examined independently. Moreover the dependence of boundary effects on flow characteristics and post-processing parameters can also be quantified. 

\begin{figure}
\begin{center}
\includegraphics[width=\textwidth]{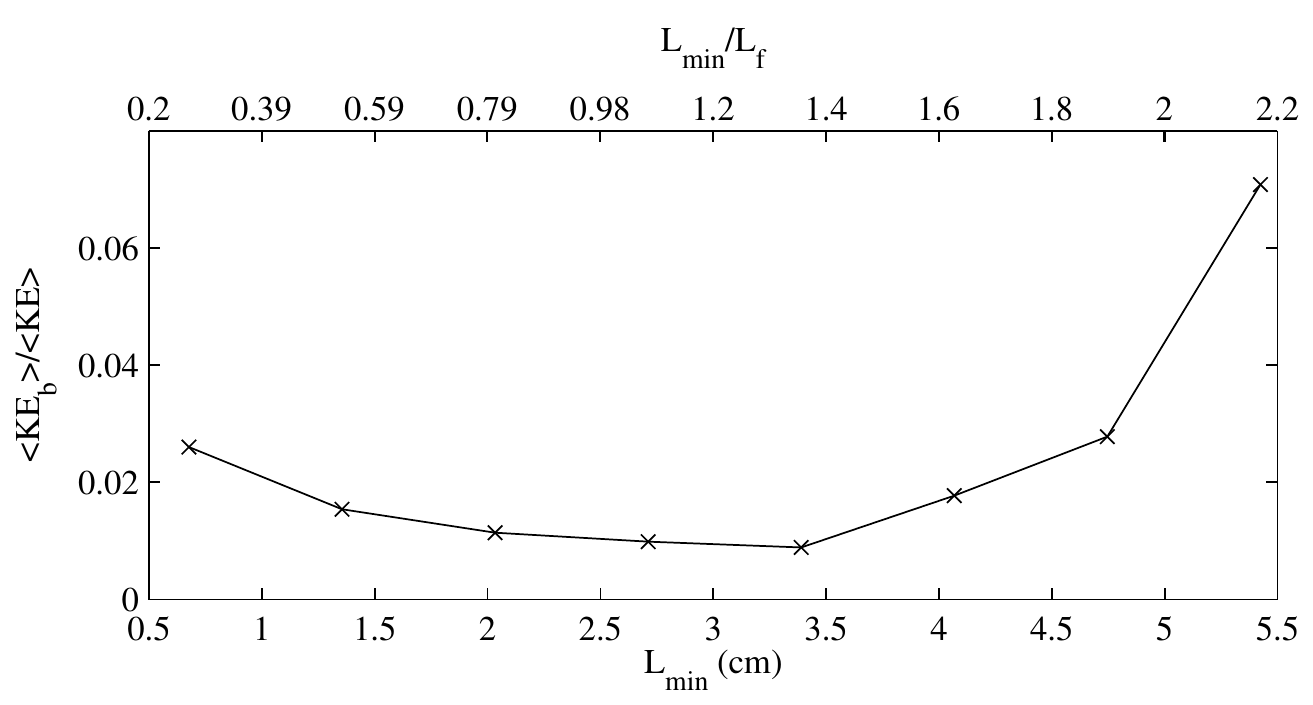} 
\caption{\label{fig:manybdry} Fraction of the total kinetic energy accounted by the boundary modes, as a function of the minimum feature size $L_{min}$, in an example frame of data with $Re=100$. Here $\left< KE_b \right>$ is the mean particle kinetic energy in the boundary modes, and $\left< KE \right>$ is the mean particle kinetic energy after reconstruction with all modes.}
\end{center}
\end{figure}

For example, the fraction of total kinetic energy associated with the boundary modes varies with $L_{min}$, as shown in figure~\ref{fig:manybdry}. When $L_{min}$ is chosen near the forcing scale $L_f$, boundary modes are relatively unimportant, accounting for less than 2\% of the total kinetic energy. As $L_{min}$ increases above the forcing scale, the boundary modes become more important, likely because interior velocity variations at scales near $L_f$ are being replaced with their near-zero average values, decreasing the contribution of the interior modes. As $L_{min}$ decreases below the forcing scale, boundary modes again become more important. The origin of this effect is less clear, but may relate to under-sampling. In the frame of data used to produce figure~\ref{fig:manybdry}, the measurement region is bounded by 183 particles, distributed around a region that is nearly rectangular and has a perimeter near 110~cm; thus their mean spacing is 0.6~cm. Though the sampling theorem applies only in the case of equal spacing, one would nonetheless expect the smallest resolvable scales to be nearly twice that length (1.2~cm), and that aliasing effects might distort measurements at smaller scales. Figure~\ref{fig:manybdry} shows the relative kinetic energy of the boundary modes beginning to increase at scales near 1.2~cm, consistent with this hypothesis.

The choice of boundaries affects the importance of the boundary modes more dramatically than the choice of $L_{min}$. By examining sub-regions of our camera's field of view, we observe that boundary modes account for a larger fraction of the total kinetic energy when the region is small than when the region is large. Since the flow is dominated by a lattice of vortices with well defined length scale, boundary modes become unimportant only in the region more than one vortex diameter from the boundary. Keeping the vortex scale constant, a decrease in the boundary size by some factor $C$ reduces the area where boundary modes are unimportant by a factor $C^2$. Thus boundary modes have greater influence in smaller measurement regions.

\begin{figure}
\begin{center}
\includegraphics{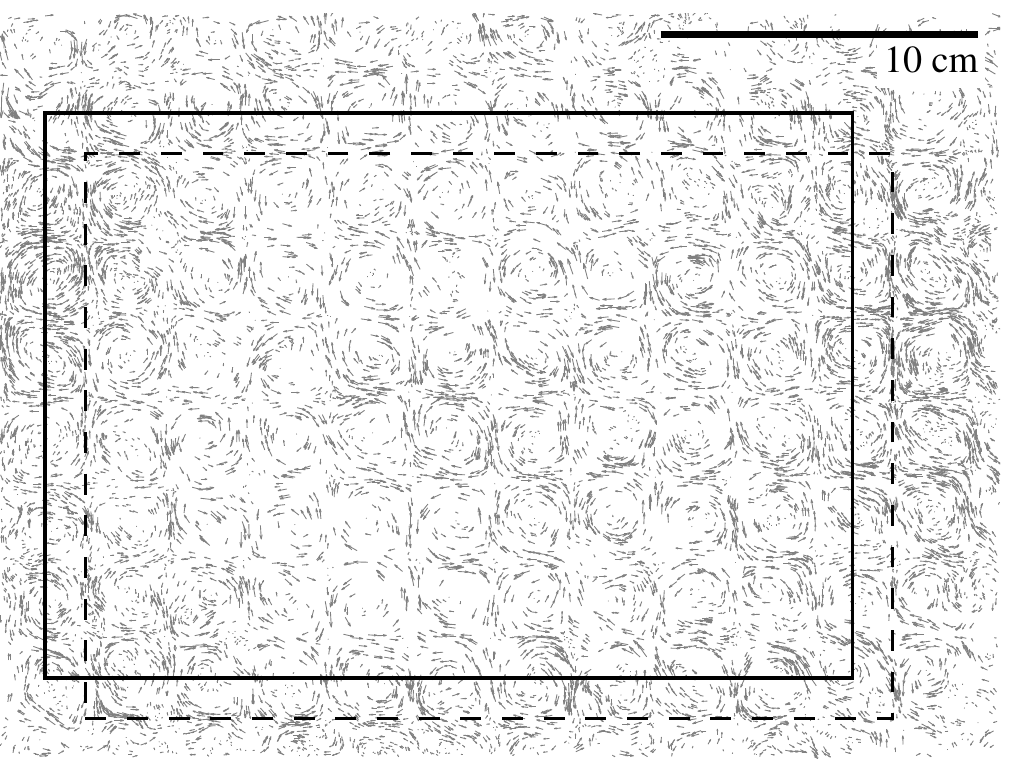} 
\caption{\label{fig:inoutregions} Regions of varying local inflow and outflow in an example velocity field as seen from above. An arrow represents the velocity of each tracked particle. The solid rectangle, cutting through vortex centers, comprises an open boundary with large local inflow and outflow. The dashed rectangle, passing near vortex separatrices, comprises an open boundary with small local inflow and outflow.}
\end{center}
\end{figure}

Changing the location of the measurement region, even while maintaining its size, can also affect the structure and scales accounted by the boundary modes. Consider the two boundaries sketched in figure~\ref{fig:inoutregions}, one drawn as a solid line and cutting through vortex centers, the other drawn as a dashed line and passing near the separatrices between vortices. Though incompressibility requires that the net outflow be zero in both cases, the local inflow and outflow (equation~\ref{eqn:outflow}) is evidently much larger for the first boundary than for the second. Accordingly, the boundary modes are more important. Reconstructing 50~frames of the same data with each boundary, we find that in the first case, boundary modes account for 18.4 $\pm$ 7.4\% of the kinetic energy, whereas in the second case, boundary modes account for just 8.1 $\pm$ 3.0\% of the kinetic energy. 

The fraction of total kinetic energy accounted by the boundary modes also varies with other parameters. Preliminarily we observe that it increases with Reynolds number. Larger Reynolds numbers may drive more of the enstrophy to scales smaller than $L_{min}$, causing the interior flow to be damped by averaging in the same way we have postulated above, for the case of large $L_{min}$.

Finally, before leaving the topic of boundary effects, we note that we have also implemented a reconstruction algorithm more closely resembling that described by~\citet{Chu:2003}. Instead of accounting for outflow at open boundaries by using the velocity potential function, it adjusts the streamfunction modes directly by solving the eigenvalue problem~\ref{eqn:eigval} with a boundary condition that depends on the outflow (instead of $\left. \Psi \right|_\Gamma = 0$). Consistent with the comments of~\citet{Lekien:2004}, however, we found that the algorithm led to problematic numerical instability. The reconstructed velocity field varied widely from frame to frame in an essentially unpredictable way, leading us to use the velocity potential function method instead. 

\section{Conclusions}
\label{sec:conclusion}

The post-processing technique described in \S\ref{sec:reconstruction} allows the multi-scale reconstruction of a velocity field using particle tracks. It requires neither interpolation nor smoothing, but automatically eliminates outliers and reduces spurious, apparent compressibility by projecting the measurements onto a set of incompressible, two-dimensional basis modes. By building a mesh from the Delaunay triangulation of the particle locations themselves, it is possible to calculate the necessary spatial gradients via finite-element techniques, without constructing an arbitrary grid. A set of boundary modes $\phi_j$ approximates the velocity potential function, accounting for local inflows and outflows at the boundary. They are constructed numerically by solving Laplace's equation. The streamfunction, which accounts for the remaining flow, is expanded in a set of interior modes $\psi_k$. For these, the eigenmodes of the Laplacian form a natural basis. In irregular measurement regions, the eigenmodes may be calculated numerically. In regions of regular shape in which analytic eigenmodes are known, they may be used instead, reducing the computational expense. For our data, we have chosen Fourier modes, since the boundaries of our effective measurement region nearly match the rectangular field of view of our camera. 

The resulting reconstructed velocity fields give direct access to scale-local properties of the flow and can be decomposed without filtering. Spatial spectra can be constructed without windowing, boundary effects can be separated from the interior flow modes, and the technique holds promise for direct calculation of enstrophy and energy transfer rates. 

This technique can be applied to any set of two-dimensional flow measurements that can be expressed as a list of local velocity measurements at particular locations over time. In its current form, it is better suited for laboratory flows with high particle density than for oceanic flows with low density, since we have avoided supplementing the measurements with interpolation or modeling. Stratified flow experiments, rotating quasi-two-dimensional experiments, and soap film experiments are natural candidates for this type of post-processing, and we expect it to have wide applicability. 

\textit{This work was supported by the U.S. National Science Foundation under Grant No.~DMR-0906245.}

\bibliographystyle{plainnat}
\bibliography{reading}

\begin{thebibliography}{26}
\providecommand{\natexlab}[1]{#1}
\providecommand{\url}[1]{\texttt{#1}}
\expandafter\ifx\csname urlstyle\endcsname\relax
  \providecommand{\doi}[1]{doi: #1}\else
  \providecommand{\doi}{doi: \begingroup \urlstyle{rm}\Url}\fi

\bibitem[Arfken and Weber(2001)]{Arfken:2001}
G.~B. Arfken and H.~J. Weber.
\newblock \emph{Mathematical Methods for Physicists}.
\newblock Harcourt Academic Press, San Diego, fifth edition, 2001.

\bibitem[Bruneau et~al.(2007)Bruneau, Fischer, and Kellay]{Bruneau:2007}
C.-H. Bruneau, P.~Fischer, and H.~Kellay.
\newblock The structures responsible for the inverse energy and the forward
  enstrophy cascades in two-dimensional turbulence.
\newblock \emph{Europhys. Lett.}, 78\penalty0 (3):\penalty0 34002, 2007.
\newblock URL \url{http://stacks.iop.org/0295-5075/78/34002}.

\bibitem[Bullard and Gellman(1954)]{Bullard:1954}
E.~Bullard and H.~Gellman.
\newblock Homogeneous dynamos and terrestrial magnetism.
\newblock \emph{Phil. Trans. R. Soc. London A}, 247\penalty0 (928):\penalty0
  213--278, November 1954.

\bibitem[Chu et~al.(2003)Chu, Ivanov, Korzhova, Margolina, and
  Melnichenko]{Chu:2003}
P.~C. Chu, L.~M. Ivanov, T.~P. Korzhova, T.~M. Margolina, and O.~V.
  Melnichenko.
\newblock Analysis of sparse and noisy ocean current data using flow
  decomposition. {P}art {I}: Theory.
\newblock \emph{J. Atmos. Ocean. Tech.}, 20\penalty0 (4):\penalty0 478--491,
  2003.

\bibitem[Clercx et~al.(2003)Clercx, van Heijst, and Zoeteweij]{Clercx:2003}
H.~J.~H. Clercx, G.~J.~F. van Heijst, and M.~L. Zoeteweij.
\newblock Quasi-two-dimensional turbulence in shallow fluid layers: The role of
  bottom friction and fluid layer depth.
\newblock \emph{Phys. Rev. E}, 67\penalty0 (6):\penalty0 066303, Jun 2003.
\newblock \doi{10.1103/PhysRevE.67.066303}.

\bibitem[Kraichnan(1967)]{Kraichnan:1967}
R.~H. Kraichnan.
\newblock Inertial ranges in two-dimensional turbulence.
\newblock \emph{Phys. Fluids}, 10:\penalty0 1417--1423, July 1967.

\bibitem[Lekien et~al.(2004)Lekien, Coulliette, Bank, and Marsden]{Lekien:2004}
F.~Lekien, C.~Coulliette, R.~Bank, and J.~Marsden.
\newblock Open-boundary modal analysis: Interpolation, extrapolation, and
  filtering.
\newblock \emph{J. Geophys. Res.}, 109:\penalty0 C12004, 12 2004.
\newblock URL \url{http://dx.doi.org/10.1029/2004JC002323}.

\bibitem[L\"{u}thi et~al.(2005)L\"{u}thi, Tsinober, and Kinzelbach]{Luthi:2005}
B.~L\"{u}thi, A.~Tsinober, and W.~Kinzelbach.
\newblock Lagrangian measurement of vorticity dynamics in turbulent flow.
\newblock \emph{J. Fluid Mech.}, 528:\penalty0 87--118, 2005.

\bibitem[Lynch(1989)]{Lynch:1989}
P.~Lynch.
\newblock Partitioning the wind in a limited domain.
\newblock \emph{Mon. Weather Rev.}, 117:\penalty0 1492, 1989.
\newblock \doi{10.1175/1520-0493(1989)117}.

\bibitem[Mathur et~al.(2007)Mathur, Haller, Peacock, Ruppert-Felsot, and
  Swinney]{Mathur:2007}
Manikandan Mathur, George Haller, Thomas Peacock, Jori~E. Ruppert-Felsot, and
  Harry~L. Swinney.
\newblock Uncovering the {L}agrangian skeleton of turbulence.
\newblock \emph{Phys. Rev. Lett.}, 98\penalty0 (14):\penalty0 144502, 2007.
\newblock \doi{10.1103/PhysRevLett.98.144502}.
\newblock URL \url{http://link.aps.org/abstract/PRL/v98/e144502}.

\bibitem[Mordant et~al.(2004)Mordant, Crawford, and Bodenschatz]{Mordant:2004a}
N.~Mordant, A.~M. Crawford, and E.~Bodenschatz.
\newblock Experimental {L}agrangian acceleration probability density function
  measurement.
\newblock \emph{Physica D}, 193:\penalty0 245--251, 2004.

\bibitem[Ouellette(2010)]{Ouellette:2010}
N.~T. Ouellette.
\newblock On the performance of predictive best-estimate particle-tracking
  algorithms.
\newblock \emph{submitted}, 2010.

\bibitem[Ouellette et~al.(2006)Ouellette, Xu, and Bodenschatz]{Ouellette:2006}
N.~T. Ouellette, H.~Xu, and E.~Bodenschatz.
\newblock A quantitative study of three-dimensional {L}agrangian particle
  tracking algorithms.
\newblock \emph{Exp. Fluids}, 40:\penalty0 301--313, February 2006.
\newblock \doi{10.1007/s00348-005-0068-7}.

\bibitem[Ouellette et~al.(2007)Ouellette, Xu, and Bodenschatz]{Ouellette:2007a}
N.~T. Ouellette, H.~Xu, and E.~Bodenschatz.
\newblock Measuring {L}agrangian statistics in intense turbulence.
\newblock In C.~Tropea, A.~L. Yarin, and J.~F. Foss, editors, \emph{Springer
  Handbook of Experimental Fluid Mechanics}. Springer-Verlag, Berlin, 2007.

\bibitem[Ouellette and Gollub(2007)]{Ouellette:2007}
Nicholas~T. Ouellette and J.~P. Gollub.
\newblock Curvature fields, topology, and the dynamics of spatiotemporal chaos.
\newblock \emph{Phys. Rev. Lett.}, 99\penalty0 (19):\penalty0 194502, 2007.
\newblock \doi{10.1103/PhysRevLett.99.194502}.
\newblock URL \url{http://link.aps.org/abstract/PRL/v99/e194502}.

\bibitem[Press et~al.(2007)Press, Teukolsky, Vetterling, and
  Flannery]{Press:2007}
W.~H. Press, S.~A. Teukolsky, W.~T. Vetterling, and B.~P. Flannery.
\newblock \emph{Numerical Recipes: The Art of Scientific Computing}.
\newblock Cambridge University Press, London, third edition, 2007.

\bibitem[Rivera and Ecke(2005)]{Rivera:2005}
M.~K. Rivera and R.~E. Ecke.
\newblock Pair dispersion and doubling time statistics in two-dimensional
  turbulence.
\newblock \emph{Phys. Rev. Lett.}, 95\penalty0 (19):\penalty0 194503, Nov 2005.
\newblock \doi{10.1103/PhysRevLett.95.194503}.

\bibitem[Rivera et~al.(2003)Rivera, Daniel, Chen, and Ecke]{Rivera:2003}
M.~K. Rivera, W.~B. Daniel, S.~Y. Chen, and R.~E. Ecke.
\newblock Energy and enstrophy transfer in decaying two-dimensional turbulence.
\newblock \emph{Phys. Rev. Lett.}, 90\penalty0 (10):\penalty0 104502, Mar 2003.
\newblock \doi{10.1103/PhysRevLett.90.104502}.

\bibitem[Rossi et~al.(2006)Rossi, Vassilicos, and Hardalupas]{Rossi:2006}
L.~Rossi, J.~C. Vassilicos, and Y.~Hardalupas.
\newblock Electromagnetically controlled multi-scale flows.
\newblock \emph{J. Fluid Mech.}, 558:\penalty0 207--242, July 2006.
\newblock \doi{10.1017/S0022112006009980}.

\bibitem[Rothstein et~al.(1999)Rothstein, Henry, and Gollub]{Rothstein:1999}
D.~Rothstein, E.~Henry, and J.~P. Gollub.
\newblock Persistent patterns in transient chaotic fluid mixing.
\newblock \emph{Nature}, 401:\penalty0 770--772, October 1999.

\bibitem[Solomon and Mezi\'c(2003)]{Solomon:2003}
T.~H. Solomon and I.~Mezi\'c.
\newblock Uniform resonant chaotic mixing in fluid flows.
\newblock \emph{Nature}, 425:\penalty0 376--380, September 2003.
\newblock \doi{10.1038/nature01993}.

\bibitem[Tabeling et~al.(1991)Tabeling, Burkhart, Cardoso, and
  Willaime]{Tabeling:1991}
P.~Tabeling, S.~Burkhart, O.~Cardoso, and H.~Willaime.
\newblock Experimental study of freely decaying two-dimensional turbulence.
\newblock \emph{Phys. Rev. Lett.}, 67\penalty0 (27):\penalty0 3772--3775, Dec
  1991.
\newblock \doi{10.1103/PhysRevLett.67.3772}.

\bibitem[Toschi and Bodenschatz(2009)]{Toschi:2009}
F.~Toschi and E.~Bodenschatz.
\newblock Lagrangian properties of particles in turbulence.
\newblock \emph{Annu. Rev. Fluid Mech.}, 41:\penalty0 375--404, 2009.

\bibitem[Tropea et~al.(2007)Tropea, Scarano, Westerweel, Cavone, Meyers, Lee,
  and Schodl]{Tropea:2007}
C.~Tropea, F.~Scarano, J.~Westerweel, A.~A. Cavone, J.~F. Meyers, J.~W. Lee,
  and R.~Schodl.
\newblock Particle-based techniques.
\newblock In C.~Tropea, J.~Foss, and A.~Yarin, editors, \emph{Springer Handbook
  of Experimental Fluid Mechanics}, pages 287--362. Springer-Verlag, Berlin,
  2007.

\bibitem[Vella and Mahadevan(2005)]{Vella:2005}
Dominic Vella and L.~Mahadevan.
\newblock The ``cheerios effect''.
\newblock \emph{American Journal of Physics}, 73\penalty0 (9):\penalty0
  817--825, 2005.
\newblock \doi{10.1119/1.1898523}.
\newblock URL \url{http://link.aip.org/link/?AJP/73/817/1}.

\bibitem[Voth et~al.(2002)Voth, Haller, and Gollub]{Voth:2002}
G.~A. Voth, G.~Haller, and J.~P. Gollub.
\newblock Experimental measurements of stretching fields in fluid mixing.
\newblock \emph{Phys. Rev. Lett.}, 88\penalty0 (25):\penalty0 254501, June
  2002.
\newblock \doi{10.1103/PhysRevLett.88.254501}.

\end{thebibliography}

\end{document}